\documentclass[fleqn,usenatbib]{mnras}

\usepackage[T1]{fontenc}

\DeclareRobustCommand{\VAN}[3]{#2}
\let\VANthebibliography\thebibliography
\def\thebibliography{\DeclareRobustCommand{\VAN}[3]{##3}\VANthebibliography}

\usepackage{graphicx}	
\usepackage{amsmath}	
\usepackage{amssymb}	
\usepackage[none]{hyphenat}
\usepackage{newtxtext,newtxmath}

\title[Radio images and the jet spreading]{
Lateral spreading effects on VLBI radio images of 
neutron star merger jets}

\author[J.J. Fernández, S. Kobayashi \& G.P. Lamb]{
Joseph John Fernández,$^{1}$\thanks{E-mail: joseph.fdez21@gmail.com}
Shiho Kobayashi$^{1}$ and
Gavin P. Lamb$^{2}$
\\
$^{1}$Astrophysics Research Institute, Liverpool John Moores University, IC2, Liverpool Science Park, 146 Brownlow Hill, Liverpool L3 5RF, UK\\
$^{2}$School of Physics and Astronomy, University of Leicester, University Road, Leicester, LE1 7RH, UK}

\date{Accepted XXX. Received YYY; in original form ZZZ}

\pubyear{2021}

\begin{document}
\label{firstpage}
\pagerange{\pageref{firstpage}--\pageref{lastpage}}
\maketitle

\begin{abstract}
Very long base interferometry (VLBI) radio images recently proved to be essential in breaking the degeneracy in the ejecta model for the neutron star merger GW170817.  We discuss the properties of synthetic radio images of merger jet afterglows by using semi-analytic models of laterally spreading or non-spreading jets. The image centroid initially moves away from the explosion point in the sky with apparent superlumianal velocity. After reaching a maximum displacement its motion is reversed. This behavior is in line with that found in full hydrodynamics simulations. We show that the evolution of the centroid shift and image size are significantly different when lateral spreading is considered. For Gaussian jet models with plausible model parameters, the morphology of the laterally spreading jet images is much closer to circular. The maximum displacement of the centroid shift and its occurrence time are smaller/earlier by a factor of a few for spreading jets. 
Our results indicate that it is crucial to include lateral spreading effects when analyzing radio images of neutron star merger jets. We also obtain the viewing angle $\theta_{\rm obs}$ by using the centroid shift of radio images  provided the ratio of the jet core size $\theta_{c}$ and $\theta_{\rm obs}$ is determined by afterglow light curves. We show that a simple method based on a point-source approximation provides reasonable angular estimates ($10-20\%$ errors at most). By taking a sample of laterally spreading structured Gaussian jets,  we obtain 
$\theta_{\rm obs} \sim 0.32$ for GW 170817, consistent with previous studies.
\end{abstract}

\begin{keywords}
Transients: gamma-ray bursts - Transients: neutron star mergers - Physical data and processes: gravitational waves - methods: numerical.
\end{keywords}



\section{Introduction}

Shortly after the detection of gravitational waves (GW) from the binary neutron star merger, GW170817, its electromagnetic (EM) counterpart was discovered in the S0 galaxy NGC4993 \citep[][etc]{coulter2017, Soares_Santos_2017}. This transient was subject to an unprecedented follow-up campaign across the EM spectrum \citep[e.g.][]{Abbott_2017_EM}. The counterpart was found to be made up of several components: a prompt short gamma-ray burst (GRB) detected $1.7s$ after the merger, a kilonova, and a broad synchrotron afterglow, first detected $9$ days post-merger at X-ray wavelengths; see \cite{margutti2020} for a review of GW170817 and \cite{metzger_2018}, \cite{burns2020} and \cite{nakar2020} for reviews of EM counterparts to GW detectable compact binary mergers.

In addition to the merger afterglow light curves, very long baseline interferometry (VLBI) radio images were obtained \citep{Mooley_2018, Ghirlanda_2019}. \cite{Mooley_2018} presented radio images at 75 and 230 days post-merger finding an image centroid displacement of $\sim 2.67\ \pm 0.2\ \textrm{mas}$ in the sky, and implying a mean apparent velocity of $\beta_{\rm app}=4.1 \pm 0.5$;
\cite{Ghirlanda_2019} confirmed this result with a radio image obtained at $207$ days post merger.
This, along with the steep post-peak afterglow decline \citep{Lamb_2018, Lamb_2019_2, Troja_2018a, Troja_2019a}, broke the degeneracy between a wide, quasi-isotropic ejecta and a 
narrow 
core-dominated jet, confirming the emission was from the latter. 

VLBI images are also important for breaking degeneracies in parameter estimation from light curves. 
\cite{nakar_piran_2020} showed that afterglow light curves observed around their peak time $T_{p}$ cannot constrain the observing angle, $\theta_{\rm obs}$, but
only determine 
the ratio of the observing angle $\theta_{\rm obs}$ to the jet opening angle, $\theta_{c}$ (or core size for a core dominated structured jet). 
This  leads to degeneracy among the parameters $\{\theta_{\rm obs},\ \theta_{c}, l, \epsilon_{B}\} $, where $l   \sim (E/nm_{p}c^{2})^{1/3}$ is the Sedov length, $n$ the ambient density, $m_{p}$ the proton mass and $\epsilon_B$ is the fraction of the shock energy in the magnetic fields. 
This degeneracy can be broken by the observation of afterglow images (the centroid shift around the peak time) or observing the light curve transition to the sub-relativistic phase \citep{nakar_piran_2020}.

Gravitational waves from compact binary mergers provide a luminosity distance $D_L$ which is independent of the cosmological distance ladder \citep{Schutz_1986, Holtz_Hughes_2005}. 
Therefore, well localized GW signals can in principle be used to estimate Hubble's constant $H_{0}$ if combined with EM redshift measurements (e.g. \citealt{Mastrogiovanni}). However the distance $D_L$ and inclination with respect to the binary plane $\theta_{\rm obs}$ are entangled in the GW strain \citep{misner, Holtz_Hughes_2005}. Without accurate GW polarization measurements to obtain $\theta_{\rm obs}$, this degeneracy leads to additional uncertainties in $D_L$.



Radio images were used in \cite{Mooley_2018,Ghirlanda_2019} to constrain  $\theta_{\rm obs}$. This data was subsequently used by \cite{Hotokezaka_2018} and \cite{wang} to constrain the Hubble constant to $H_{0}=68.9^{+4.7}_{-4.6} \textrm{km}\ \textrm{s}^{-1}\textrm{Mpc}^{-1}$ and $H_{0}=69.5^{+4}_{-4} \textrm{km}\ \textrm{s}^{-1}\textrm{Mpc}^{-1}$ respectively. The semi-analytic afterglow models used in these works were limited to non-spreading jets. 

GRB and merger jets have been studied extensively in the literature. Semi-analytic models based on non-spreading jets have been successful in the study of afterglow light curves. However, jets are expected to laterally expand at late times. As lateral spreading is most significant at late times, after the light curve peaks, this effect is often not included in semi-analytic modelling (e.g. \citealt{Salafia_2019, Ghirlanda_2019, Beniamini_2020}). However, lateral spreading hastens the light curve peak. In the case of afterglows observed off-axis, it also steepens the rise of the light curve \citep{lamb2021_2}. As we show in this work, it can also modify the properties of radio images. Therefore, inclusion of spreading effects is crucial to study the light curves and images. In this paper the properties of synthetic images of both non-spreading and laterally spreading jets are studied. In \S\ref{sec:numerical} we describe the numerical method used to model the images. In \S\ref{sec:imagediscussion} we consider laterally spreading/non-spreading jets to evaluate the afterglow images. We demonstrate the importance of including lateral spreading in semi-analytic image calculations. In \S\ref{sec:degendis} the results are applied to a GW17017-like system.  Summary conclusions are given in \S\ref{sec:conclusions}.

\section{Numerical model} \label{sec:numerical}

\subsection{Jet dynamics}  \label{sec:dynamicssection}
We consider a spherical shell, with energy $E$, mass $M=E/\Gamma_{0} c^{2}$ expanding adiabatically with an initial Lorentz factor $\Gamma_{0}$ into a cold and uniform circumburst medium (CBM) of particle density $n$. By considering conservation of the stress-energy tensor it can be shown that the Lorentz factor $\Gamma$ of the shell evolves with swept-up mass $m$ as \citep{peer2012}
\begin{eqnarray} \label{eq:peer}
\frac{d\Gamma}{dm} = -\frac{\hat{\gamma}(\Gamma^{2}-1)-(\hat{\gamma}-1)\Gamma\beta^{2}}{M + m[2\hat{\gamma}\Gamma -(\hat{\gamma}-1)(1+\Gamma^{-2})]},
\end{eqnarray}
where $\beta=\sqrt{1-\Gamma^{-2}}$ is the velocity of the shock in units of $c$ and $\hat{\gamma}$ is the adiabatic index of the fluid (see \citealt{peer2012} and \citealt{Lamb_2018} for details). Because of the relativistic beaming effect, the radiation from a jet can be described by a spherical model with an isotropic explosion energy $E$. The actual energy in the jet with a solid angle $\Omega$ is given by $(\Omega/4\pi)E$, and as the bulk of the system is causally disconnected from its edge the spherical model holds while $\Gamma>>1/\theta_{j}$, where $\theta_{j}$ is the half opening angle of the jet. As the jet decelerates, information about pressure gradients, transported by sound waves, can reach the edges, forcing them to spread laterally. To model this spreading effect we follow \cite{granotpiran12}. The opening angle $\theta_{j}$ is constant for $R<R_{d}$, where $R_{d}= l/\Gamma_{0}^{2}$ is the deceleration radius. The shell has decelerated to $\Gamma\approx \Gamma_{0}/2$ at this radius. For $R>R_{d}$, we approximate the lateral spreading of a jet (spreading cases) as\footnote{This corresponds to the $a=1$ case in \cite{granotpiran12}. In their recipe, this approximation is valid both in the relativistic and Newtonian regimes.}
\begin{eqnarray} \label{eq:lateralspreading}
\frac{d\theta_{j}}{dln\ R} = \frac{1}{\Gamma^{2}\theta_{j}}.
\end{eqnarray}
Since initially $\Gamma \theta_{j}>>1$ is satisfied lateral spreading is negligible. After the jet break when $\Gamma \sim 1/\theta_{j}$ the opening angle can grow quickly (e.g. \citealt{granotpiran12}). 
The swept-up mass evolves as,
\begin{eqnarray} \label{eq:sweptmass}
\frac{dm}{dR} = 2\pi n m_{p} \left[(1-\cos{\theta_{j}})R^{2} + \frac{1}{3}\sin{\theta_{j}}R^{3}\frac{d\theta_{j}}{dR}\right],
\end{eqnarray}
where it is assumed that the jet front is a conical section of a sphere. For the non-spreading case equation \ref{eq:sweptmass} gives $m=(\Omega/3) nm_{p}R^{3}$.

In the thin-shell approximation, photons are emitted from the blast wave surface. The arrival time $T$ (i.e. observer time) of photons emitted from a fluid element at a radius $R$ is given by (e.g. \cite{Resmi2018, Lamb_2018, Lu_2020})
\begin{eqnarray}\label{eq:obsTime}
T(R, \alpha) = \int_{0}^{R} \frac{dR}{\beta c} - \frac{R}{c} \cos{\alpha},
\end{eqnarray}
where $\alpha$ is the angle between the emitter's direction of motion and the observer line-of-sight (LOS).

\begin{figure} 
\centering
\includegraphics[width=0.9\columnwidth, height=0.30\textheight]{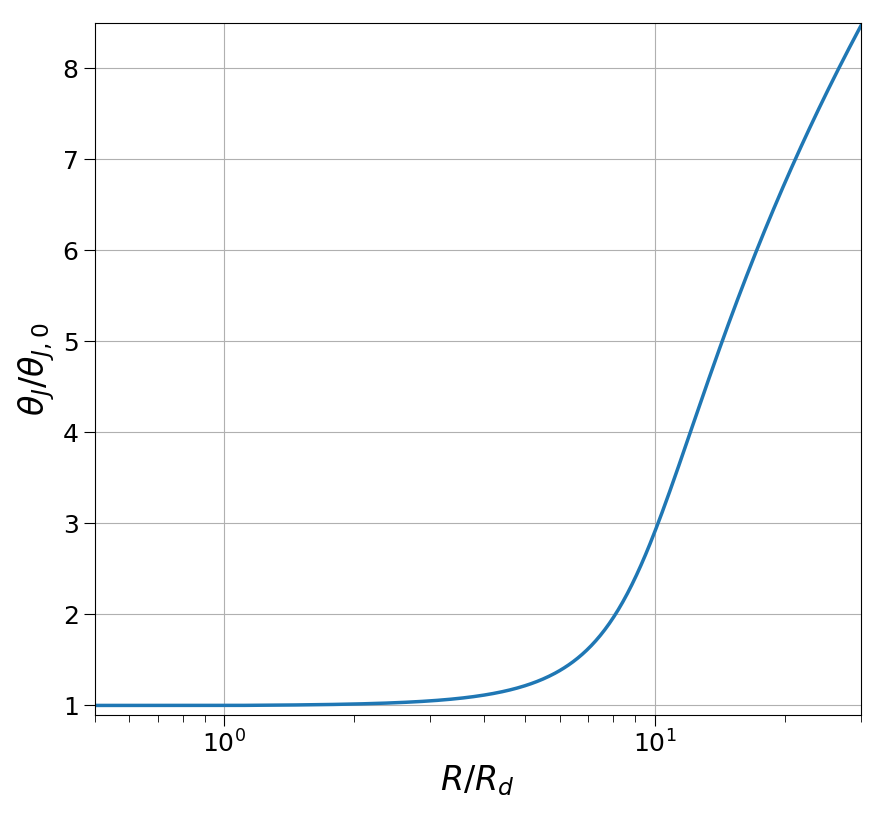}
\caption{Evolution of the opening angle $\theta_{j}$ in units of its initial value $\theta_{j,0}$. 
}
\label{fig:thetaEv}
\end{figure}
\begin{figure} 
\centering
\includegraphics[width=1.0\columnwidth, height=0.30\textheight]{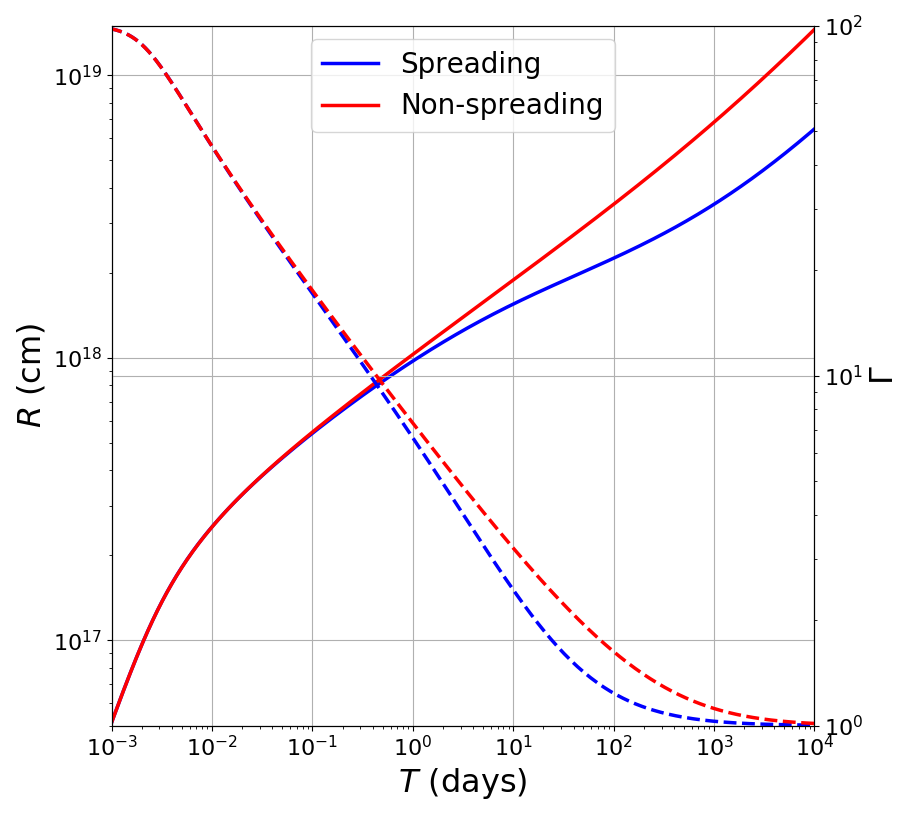}
\caption{Evolution of a top-hat jet radius $R$ (solid lines, left vertical axis) and Lorentz factor $\Gamma$ (dashed lines, right vertical-axis), as a function of $T$ 
where $T$ is evaluated for  $\alpha=0$.
Blue and red lines indicate the spreading and non-spreading case, respectively.}
\label{fig:doublepanel}
\end{figure}

To illustrate the typical jet evolution, we consider a jet with parameters $E=5\times 10^{51}\ \textrm{ergs}$, $\Gamma_{0}=100$, initial half-opening angle $\theta_{j,0}=10^{\circ}$ and $n=10^{-2}\ \textrm{cm}^{-3}$; throughout this section the on-axis case is assumed ($\theta_{\rm obs}=0$). 
The dynamical equations \ref{eq:peer}, \ref{eq:lateralspreading} and \ref{eq:sweptmass} are solved using a fourth-order Runge-Kutta scheme. 
The deceleration radius for these parameters is $R_{d}=2\times 10^{17}$\,cm, which corresponds to an on-axis deceleration time $T_{d}\approx 300$\,s\,$\sim 4\times 10^{-3}$ days. Figure \ref{fig:thetaEv} shows the evolution of the opening angle $\theta_{j}$ in units of its initial value $\theta_{j,0}$ as a function of the radius of the shell. 
The opening angle initially grows slowly and by the jet break, when $\Gamma\sim 1/\theta_{j}$ when the jet has expanded to a radius $R\approx 5 R_{d}$ (or equivalently at at a time $T\approx 1.3$ days for an ox-axis observer), the opening angle has grown by a factor of $1.25$. 
As the amount of swept-up mass is larger in the spreading case, the spreading jet decelerates faster. This is illustrated in figure \ref{fig:doublepanel}, which shows the evolution of the $R$ (solid curves) and $\Gamma$ (dashed curves) as a function of $T$ where $T$ is evaluated for $\alpha=0$.  
The blue lines correspond to the laterally spreading case, and the red lines correspond to the non-spreading case. 
The non-spreading jet begins to lead the spreading jet after $T\sim 1$ days, and by $T=75$ days has a radius which $50\%$ larger than the spreading jet. The initial evolution of the Lorentz factor $\Gamma$ in both cases is similar.
After the initial coasting phase $\Gamma$ decreases as $\Gamma\propto T^{-3/8}$, as is expected for a relativistic 
blast wave. 
After $T\sim 0.5$ days, the evolution of the two systems separate. 
The spreading jet presents a steeper decay until $T\approx 1000$ days, when its Lorentz factor begins to flatten as the system approaches the sub-relativistic regime.   

\subsection{Synchrotron spectrum}\label{sec:syncform}

The shock accelerates the CBM electrons to a power-law distribution $N(\gamma_{e})d\gamma_{e}\propto \gamma_{e}^{-p}d\gamma_{e}$, for $\gamma_{e}>\gamma_{m}$. Here $\epsilon_{e}$ is the fraction of shock energy given to the electrons, $\gamma_{m}=\epsilon_{e} (p-2)(p-1)^{-1} (m_{p}/m_{e}) (\Gamma-1)$ is the minimum electron Lorentz factor and $m_{e}$ are the mass of the electron. 
The magnetic field strength behind the shock  is $B=\{ 8\pi \epsilon_{B} n m_{p}c^{2}(4\Gamma + 3)(\Gamma -1) \}^{1/2}$. 

Relativistic electrons emit synchrotron radiation as they gyrate around the magnetic field lines. The characteristic synchrotron frequency is $\nu_{\rm sync}(\gamma_{e}) = (qB\gamma_{e}^{2}/2\pi m_{e}c) \delta$, where $q$ the electron charge and $\delta=\Gamma^{-1}(1-\beta \cos{\alpha})^{-1}$ is the Doppler factor. 

The synchrotron spectrum is described by a broken power law with two break frequencies, $\nu_{m}=\nu_{sync}(\gamma_{m})$ and $\nu_{c}=\nu_{sync}(\gamma_{c})$, where $\gamma_{c}=(6\pi m_{e}c)/(\sigma_{T} \Gamma B^{2}T(\alpha=0))$ (e.g. \citealt{spn98}). 
The spectrum is given by
\begin{eqnarray}
    \frac{F_{\nu}}{F_{\nu, {\rm max}}} = 
    \begin{cases}
            (\nu/\nu_{m})^{1/3}  & \nu_{m}>\nu, \\
            (\nu/\nu_{m})^{-(p-1)/2} &  \nu_{c}>\nu>\nu_{m}, \\
            (\nu_{c}/\nu_{m})^{-(p-1)/2}(\nu/\nu_{c})^{-p/2}  &  \nu>\nu_{c}, \\
    \end{cases}
\end{eqnarray}
where the peak flux $F_{\nu,{\rm max}}=\delta^{3}(N_{e} \sigma_{T}m_{e}c^{2}B)(12\pi q D^{2}_{L})^{-1}$. Here $N_{e}$ is the number of emitting electrons and $D_{L}$ the luminosity distance to the source.

\subsection{Discretization of the system} \label{subsec:discret}

We consider an axisymmetric jet, and assume that all jet and shocked ambient material is confined to an infinitely thin region (this is known as the \textit{thin-shell} approximation). 
The jet is initially the polar region of a sphere with half opening angle $\theta_{j}$. 



For numerical purposes the jet is divided into a single central spherical cap and $n-1$ rings centered on the jet axis. The central spherical cap is labelled with k=0, and the concentric rings by $k=1,\ 2,...,\ n-1$. The spherical cap has an opening angle $\theta_{b,1}$, and the $k$-th ring is bounded by two concentric cirlces on the sphere with $\theta_{b,k}$ and $\theta_{b,k+1}$ given by
\begin{eqnarray} 
\theta_{b, k}=2\sin^{-1}\left(\frac{k}{n}\sin{\frac{\theta_{j}}{2}}\right).
\end{eqnarray}
The spherical cap is regarded as a single region, and the $k$-th ring is divided in the azimuthal direction $\phi$ into $2k+1$ equal size regions with boundaries $\phi_{kl}=2\pi l/(2k+1)$, where $l=0,\ 1,...\ k$. This division yields a total of $\sum_{k=0}^{n-1} (2k+1)=n^{2}$ regions each subtending a solid angle of $2\pi(1-\cos\theta_{j})/n^{2}$ \citep{BECKERS2012275}. Having divided the jet, the positions of the regions are defined by the radius $R$ and the angles $(\theta_{k}, \phi_{kl})$, where $\theta_{k}=(\theta_{b,k}+\theta_{b,k-1})/2$, $\phi_{kl} = (\phi_{kl} + \phi_{kl-1})/2$. The coordinate vector $\mathbf{r}_{kl} = (x_{kl}, y_{kl}, z_{kl})$ of the fluid element $kl$ has components
\begin{eqnarray}
\begin{cases}
x_{kl} = R\sin{\theta_{k}}\cos{\phi_{kl}}, \\
y_{kl} = R\sin{\theta_{k}}\sin{\phi_{kl}}, \\
z_{kl} = R\cos{\theta_{k}}.
\end{cases}
\end{eqnarray}
Given a set of initial conditions, 
only the radial component $R$ will evolve for a non-spreading jet, whereas, for a laterally spreading jet both $R$ and $\theta_k$ will evolve as we discuss in section \ref{sec:imagediscussion}.

\subsection{Construction of light curves and synthetic images}

To obtain light curves we use the procedure outlined in \cite{lamb_17} and \cite{Lamb_2018}. We take a Cartesian coordinate system in which the jet propagates in the $z$-direction and the observer is in the $yz$-plane.  
The LOS forms an angle $\theta_{\rm obs}$ with the $z$-axis. 
The direction along the LOS is defined by the unit vector $\hat{\mathbf{n}}_{\rm obs}=\sin{\theta_{\rm obs}} \hat{\mathbf{y}} +\cos{\theta_{\rm obs}} \hat{\mathbf{z}}$. The angle between the LOS and the direction of motion of the fluid element $\hat{\mathbf{r}}_{kl}=\mathbf{r}_{kl}/R_{kl}$, for a non-spreading jet, is given by $\alpha_{\rm kl}=\cos^{-1}{(\hat{\mathbf{r}}_{kl}\cdot \hat{\mathbf{n}}_{\rm obs})}$.  
The contribution of this cell to the light curve at time $T$ is obtained by inverting equation \ref{eq:obsTime} to obtain the observed radius $R_{kl}(T)$, which determines the emission with the formalism detailed in section \ref{sec:syncform}. 
For a laterally spreading jet the fluid element also has a sideways expansion velocity. 
However, the lateral velocity is much smaller than the radial velocity so the Doppler factor is evaluated by using the direction of radial motion only. 
The light curve is obtained by adding up the contribution of each individual fluid element, i.e. $F_{\nu}(T)=\sum_{k,l} F_{\nu,kl}(T)$.

The imaging plane is perpendicular to the LOS of the distant observer. 
Two mutually perpendicular directions in this plane are given by the basis vectors $\hat{\tilde{\mathbf{x}}}=\sin{\theta_{\rm obs}}\hat{\mathbf{z}} - \cos{\theta_{\rm obs}}\hat{\mathbf{x}}$, $\hat{\tilde{\mathbf{y}}}=\hat{\mathbf{x}}$, where the tildes indicate vectors in the imaging plane, as seen by an off-axis observer. 
This basis is chosen so that the principal jet moves in the positive $\tilde{x}$-direction in the imaging plane. Having defined the unit vectors in the imaging plane, the coordinates of the fluid elements in the image are given by $\tilde{x}_{kl}=\mathbf{r}_{kl}\cdot    \hat{\tilde{\mathbf{x}}}$, $\tilde{y}_{kl}=\mathbf{r}_{kl}\cdot \hat{\tilde{\mathbf{y}}}$. The specific intensity or brightness of a given fluid element is obtained from the spectral flux by considering the solid angle subtended by the emitter in the sky.


\subsection{Jet structures} \label{subsec:jetstruc}
Full hydrodynamics simulations show that when a merger jet needs to drill through surrounding ejecta, the emerging jet has a specific structure in energy and Lorentz factor distributions \citep{De_Colle_2012, Xie_2018_18, Gottlieb_2020, lorenzo}. These structures affect the time evolution and shape of afterglow light curves (e.g. \citealt{Granot_2002_3, Wei_2003, Zhang_2002, Rossi_2004, Granot_Kumar_2003, Salafia:2015vla, lamb_17, Beniamini_2020, Takahashi}). 

Here we consider a Gaussian jet model characterized by a core with semi-opening angle $\theta_{c}$ within which most of the energy is contained. 
The energy per unit solid angle, $\epsilon(\theta)$, and the initial Lorentz factor, $\Gamma_0(\theta)$, distributions are,

\begin{eqnarray} \label{eq:gaussianjet}
\begin{cases}
        \epsilon(\theta)\ \  = \epsilon_{c}e^{-\theta^{2}
        /\zeta_{1} \theta_{c}^{2}} \\
        \Gamma_{0}(\theta) = 1+(\Gamma_{c}-1)e^{-\theta^{2}/\zeta_{2}\theta_{c}^{2}},
\end{cases}
\end{eqnarray}
where the values $\zeta_{1}=1$, $\zeta_{2}=2$ are assumed \citep[e.g.][]{Resmi2018, Lamb_2018_2, Lamb_2019_2}.  
For these coefficients the deceleration radius $R_{d}$ does not depend on $\theta$. 
The structure is imposed as initial conditions for the dynamics.

\subsection{Implementation of lateral spreading for structured jets}
From equation \ref{eq:lateralspreading} it can be seen that the degree of lateral spreading depends on the Lorentz factor of the jet and that spreading becomes significant when $\Gamma\sim 1/\theta_{j}$. 
To include this effect in the dynamics of a top-hat jet, equation \ref{eq:lateralspreading} is simply applied for the edges and the spreading is applied to each fluid element accordingly.

For structured jets the implementation is slightly more complex. The initial conditions are given as a function of $\theta$ by equations \ref{eq:gaussianjet}, and the discretization described above divides the jet into rings of constant $\epsilon(\theta_{k}),\ \Gamma_{0}(\theta_{k})$. Each of these rings has slightly different dynamical evolution, therefore, we assume each ring to be part of a top hat with initial opening angle $\theta_{b,k+1}$. Neglecting the interaction between rings, the dynamical evolution of each ring, including the spreading effect, is approximated by using the top-hat jet model. 

Similar approaches have been considered in previous studies, including \cite{Lamb_2018} and \cite{Ryan_2020}. While more flexible and significantly less costly to run, the semi-analytic calculations of laterally spreading jet dynamics have limitations. 
First, this approximation ignores the gradual pressure gradient and instead its assumes a sharp density gradient for each ring. Thus, it overestimates the spreading of each part of the jet. Second, the obtained structure after spreading starts is inconsistent in the sense that different initial rings occupy the same region in space and are decelerated independently by the same collected mass. This approximation has not been fully tested for a structured jet (unlike top-hat jet where these problems do not exist). Although a detailed comparison with full hydrodynamic simulation results is needed to quantify errors in this approximation, considering the overestimates of the jet spreading, the real centroid shift of jet images (i.e. the full hydrodynamic simulation results) might take an intermediate value of those obtained through our two approximations.

\section{Synthetic radio images of spreading or non-spreading jets} \label{sec:imagediscussion}

Recently, 2D (e.g. \citealt{granot_2018, Zrake_2018}) and 3D hydrodynamics models (\citealt{Nakar_2018_2}) have been used to obtain synthetic images in the context of the NS merger event GW170817. \cite{granot_2018} present radio images for a uniform ISM and different wind-like profiles. They find that the image centroid is initially dominated by the principal jet contribution, but when the counter jet becomes visible this component becomes dominant rather quickly. \cite{Zrake_2018} compare the evolution from successfully launched, anisotropic jets and choked jets, which give rise to a quasi-spherical explosion. They obtain the evolution of the image centroid and width, and find that, for GW170817-like events, the latter may be used to distinguish between the choked and successful jet scenarios.

Semi-analytic models have been used extensively for light curve calculation, both for simple top-hat jets and diverse structured jets. Lateral spreading is generally introduced in semi-analytic models by assuming a simplified model, such as the formalism presented in \cite{granotpiran12}. 
\cite{Lamb_2018} and \cite{Ryan_2020} take similar approaches, modelling this feature as sound-speed expansion of the jet edges. However, the application of semi-analytic models to imaging in the literature is generally limited to non-spreading jets. Semi-analytic imaging was presented in \cite{Gill_2018} for Gaussian and power law structured, non-spreading jets; see also \cite{Ghirlanda_2019}. 
In \cite{Lu_2020}, a semi-analytic effective 1-D formalism is presented for which lateral spreading is derived by considering momentum conservation and pressure gradients (in this paper the authors show the evolution of the numerical grid points, but explicit imaging is not provided). \cite{dufflas2018} has also given detailed studies of the spreading process.

Once the solid angle of the jet increases significantly, the jet decelerates faster. The effect of lateral spreading is not substantial significant in the rising part of the light curves, however, the light curve, where $\nu<\nu_m$, peaks earlier when lateral spreading is included. 
The decay index after the jet break, or peak for an off-axis observer, depends on the lateral spreading. 

In this paper the synthetic radio images of laterally spreading jets are obtained using the semi-analytic model described in the previous section. The deceleration of the jet caused by mass build-up on the shock shell is governed by equation \ref{eq:peer}, the swept-up mass is given by equation \ref{eq:sweptmass}, and the evolution of the opening angle is given by equation \ref{eq:lateralspreading}. 

Radio imaging has been proposed as a tool to break the degeneracy between radial and angular structured jets \citep{Gill_2018} or between isotropic and jet-like ejecta \citep{Mooley_2018}. In particular, obtaining the image centroid, defined as the surface brightness-weighted centre of the image,
\begin{eqnarray}
\tilde{x}_{c} = \frac{1}{\int I_{\nu} d\tilde{x}'d\tilde{y}'}\int \tilde{x}' I_{\nu} d\tilde{x}'d\tilde{y}',
\end{eqnarray}
was key for this purpose. For spherical blast waves the centroid does not move i.e. $\tilde{x}_{c}=0$. This is also the case for jets observed exactly on axis (when the LOS runs exactly along the jet axis). For jets observed off-axis, at early times $\tilde{x}_{c}$ moves in the principal jet direction. For relativistic jets we would observe superluminal motion of the jet on the sky if the viewing angle is small.



\begin{figure} 
\centering
\includegraphics[width=0.9\columnwidth, height=0.6\textheight]{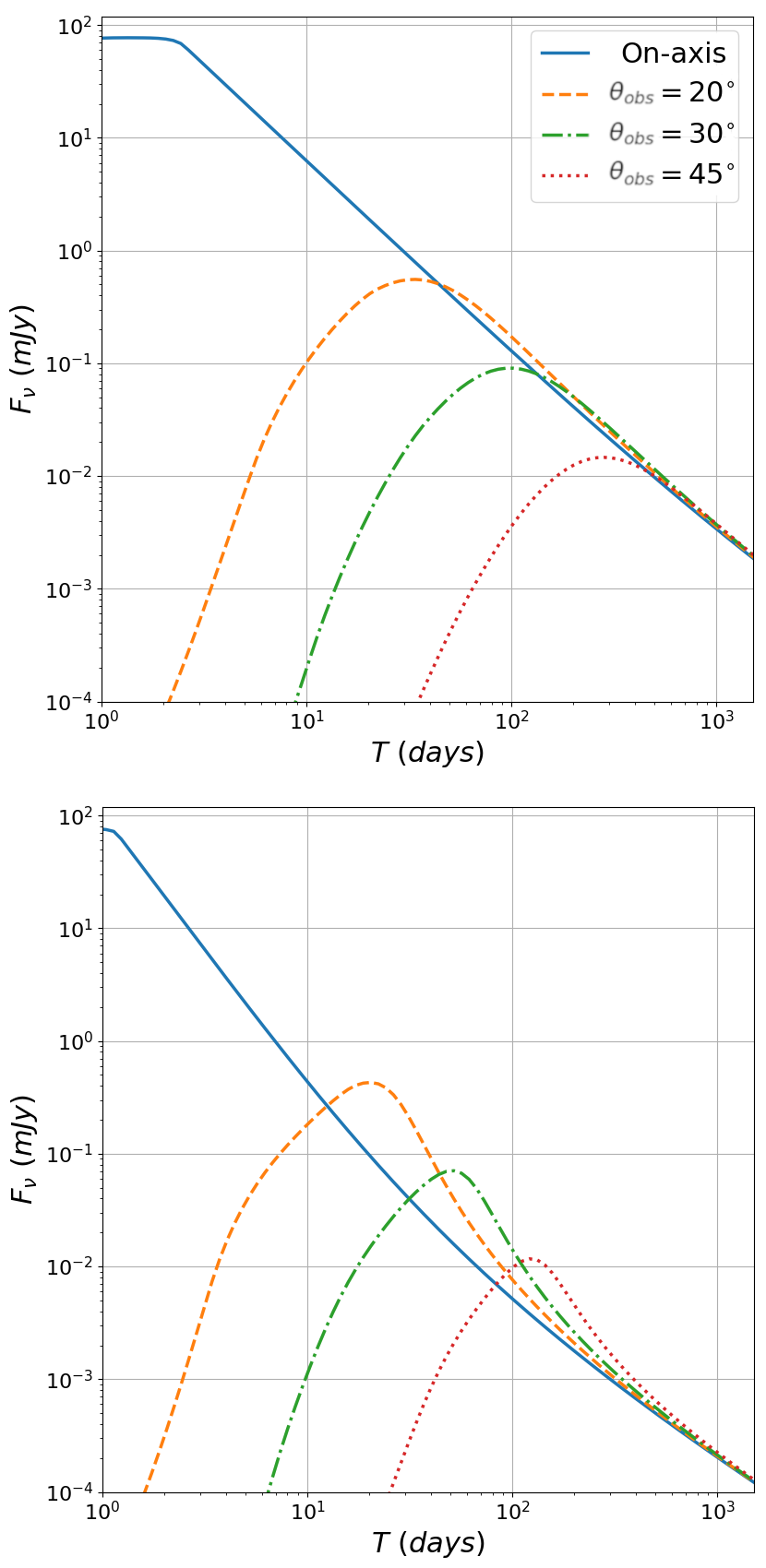}
\caption{Light curves for a Gaussian jet obtained at frequency $\nu=5\ GHz$ for a non-spreading jet case (top panel) and a case which includes lateral spreading (bottom panel). The light curves correspond to an on-axis observer (solid blue lines), and observers at $\theta_{\rm obs}=20^{\circ}$ (orange dashed lines), $\theta_{\rm obs}=30^{\circ}$ (green dashed-dotted lines) and $\theta_{\rm obs}=45^{\circ}$ (red dotted lines).}
\label{fig:light curves}
\end{figure}

To illustrate the spreading effects on light curves and images, consider a Gaussian structured with jet $E_{c}=4\pi\epsilon_{c}=5\times10^{51}\ \textrm{ergs}$, $\Gamma_{c}=100$, $n=10^{-2} \textrm{cm}^{-3}$, $\theta_{j,0}=10^{\circ}$ and $\theta_{c}=3^{\circ}$, and the microscopic parameters $p=2.16$, $\epsilon_{e}=0.1$ and $\epsilon_{B}=0.01$  (e.g. \citealt{lamb_17, Zrake_2018}). We consider a luminosity distance to the observer of $D_{L}=41.3\ \textrm{Mpc}$.
Figure \ref{fig:light curves} shows the light curves obtained for $\theta_{\rm obs}=0^{\circ},\  20^{\circ}\ , 30^{\circ} $ and $45^{\circ}$. 
The on-axis light curves are very similar for the non-spreading (top panel) and spreading cases (bottom panel). 
For the off-axis cases, the light curves obtained for the laterally spreading jet show a slightly faster rise, an earlier peak time, and a faster post-peak decay. 
The earlier brightening is due to more mass being swept-up at smaller radii when compared to the non-spreading case. 
The steeper decay after peak time is again due to lateral spreading transporting energy away from the core \citep{Lu_2020}. The peak times for the non-spreading are $T_{p}\approx 1.4, \  33,\  97$ and $284$ days, and for the spreading-jet $T_{p}\approx 0.8,\ 20,\ 51$ and $122$ days (for $\theta_{\rm obs}=0^{\circ},\  20^{\circ}\ , 30^{\circ} $ and $45^{\circ}$ respectively).

\begin{figure*} 
\centering
\includegraphics[width=1.6\columnwidth, height=0.9\textheight]{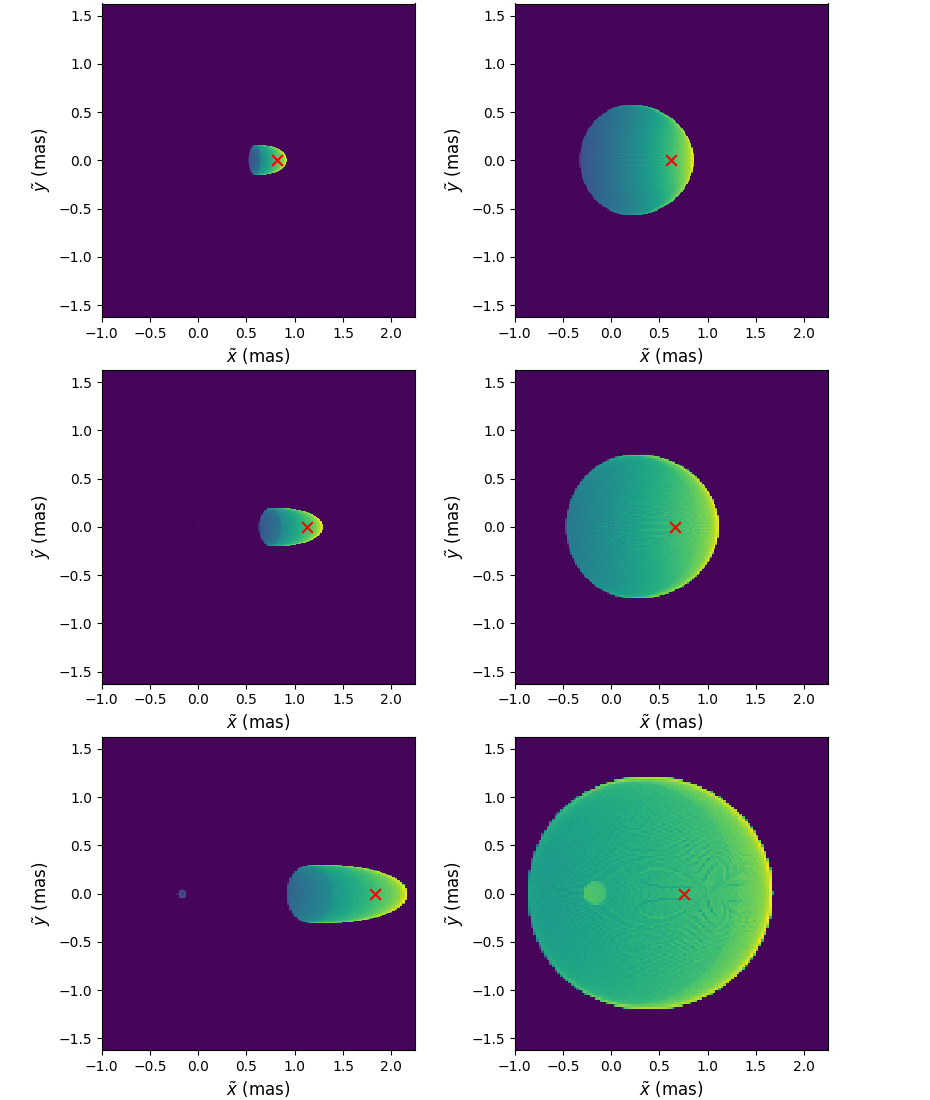}
\caption{Radio images ($\nu=5\ GHz$) at times 44, 75 and 230 days (top, middle and bottom panels respectiveley) after explosion for an observer with line of sight at $\theta_{\rm obs}=20^{\circ}$ from the jet axis and a distance of $D=41.3$ Mpc from the source. The images are normalized to the maximum brightness $I_{\nu,{\rm max}}$ in each frame, with a background threshold set at $0.01 I_{\nu,{\rm max}}$. The red crosses mark the position of the image centroid. The non-spreading case is shown in the left column and the spreading case in the right column.}
\label{fig:images20}
\end{figure*}

Figure \ref{fig:images20} shows synthetic radio images of the afterglows for  $\theta_{\rm obs}=20^{\circ}$ from the jet axis, at times $T=44, \ 75$ and $230$ days after the initial explosion. 
The surface brightness in each image is normalized to $I_{\nu,{\rm max}}$, with the colour map covering the range between $0.01 I_{\nu,{\rm max}}$ and $I_{\nu,{\rm max}}$. 
The red crosses indicate the position of the centroid in each image. 
The counter-jet is visible in the images at $T=230$ days. In the spreading case it appears a brightness excess, as the projection of the principal and counter-jets overlap in the imaging plane.

In the non-spreading case (left column) the principal jet always moves in the positive $\tilde{x}$-direction, leaving the explosion origin behind ($\tilde{x}=0$) in the imaging plane. The images resemble a sliced ellipsoid, and present a gradual decrease in brightness from front (right-most edge) to back (left-most edge). A ring develops in the back of the image, the edges of which correspond to the wings of the jet. The bright leading section corresponds to the centre of the jet. The extension along the $\tilde{x}$-direction is around $2-4$ times that along $\tilde{y}$. In contrast, the morphology of the laterally spreading jet images is much closer to circular, due to the excess (reduced) growth along $\tilde{y}$ ($\tilde{x}$). As before the images present a dim ring which encloses a brighter region corresponding again to central, more energetic fluid elements. Similar morphology is found at higher inclinations $\theta_{\rm obs}=30^{\circ},\ 45^{\circ}$. In addition, lateral spreading also causes the fluid elements to rotate around the origin of the imaging plane. When the jet-opening angle grows to the observing angle, $\theta_{j}=\theta_{\rm obs}$, the apparent motion of its outermost components is in the negative $\tilde{x}$- direction. This gives rise to the features described in following sections, such as the centroid motion reversal.

\subsection{Evolution of the centroid of jet images}
\begin{figure} 
\centering
\includegraphics[width=0.9\columnwidth, height=0.3\textheight]{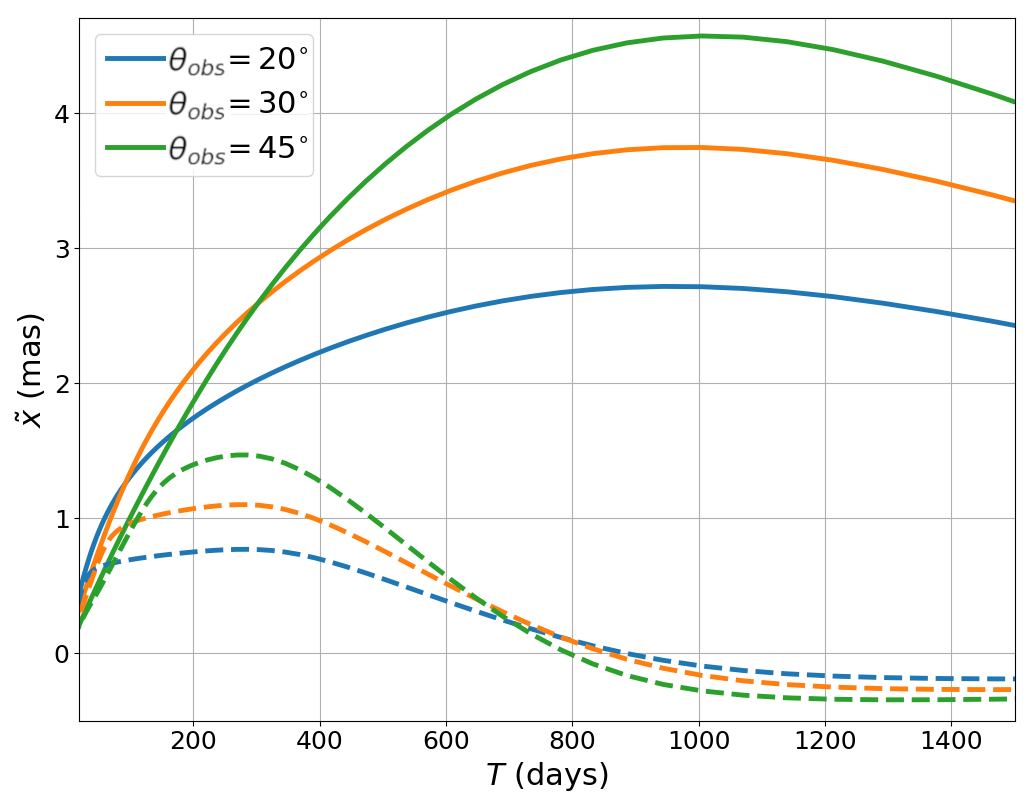}
\caption{Evolution if the image centroid position $\tilde{x}_{c}$ as a function of observer time $T$ and observing angle $\theta_{\rm obs}$ (blue: $\theta_{\rm obs}=20^{\circ}$, orange: $\theta_{\rm obs}=30^{\circ}$, green: $\theta_{\rm obs}=45^{\circ}$). The solid (dashed) lines are for the non-spreading (spreading) case.}
\label{fig:jetcentroids}
\end{figure}

The spreading and non-spreading cases result in significantly different evolutions for the centroid shifts. In figure \ref{fig:jetcentroids} the evolution of the image centroid is shown for observing angles $\theta_{\rm obs}=20^{\circ},\ 30^{\circ}$ and $45^{\circ}$. 
The centroid quickly moves away from the explosion point in the sky. 
At very late times ($T \sim 1000$ days in all three cases for the non-spreading jet), the centroid reverses its motion. 
For the laterally spreading jet the early evolution is similar, but the maximum centroid displacement is smaller and is reached much earlier, at $T\sim 340,\ 335$ and $330$ days for $\theta_{\rm obs}=20^{\circ},\ 30^{\circ}$ and  $45^{\circ}$ respectively. 
By these times, the light curve flux has decreased by a factor of $\sim 180,\ 45$ and $7$ for the spreading jet compared to the peak flux. The centroid position continues to move backwards and crosses $\tilde{x}=0$. It then reverses once more and asymptotically approaches $\tilde{x}=0$ after several thousand days. Similar centroid behavior was found using 2D hydrodynamics simulations in \cite{granot_2018} (see their figure 6).  

\begin{figure} 
\centering
\includegraphics[width=0.9\columnwidth, height=0.6\textheight]{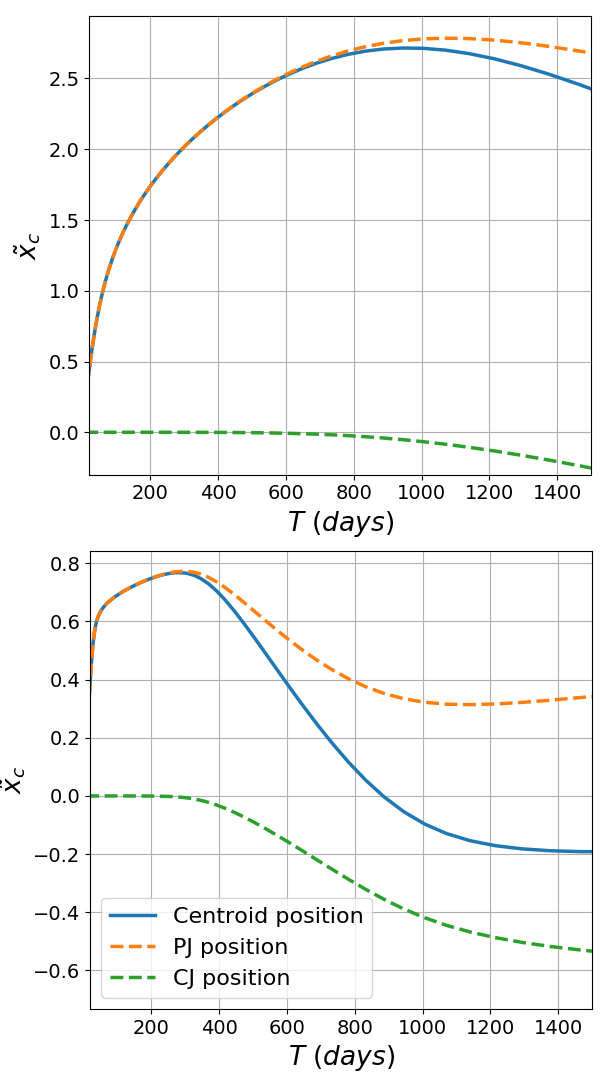}
\caption{Position of the image centroid $\tilde{x}_{c}$ (solid blue curve) as a function of time $T$ for the non-spreading (top panel) and spreading case (bottom panel) for $\theta_{\rm obs}=20^{\circ}$. The brightness-weighted contribution of the principal jet (dashed orange curves) and counter jet (dashed green curves) are also shown.}
\label{fig:jetcentroidscontr}
\end{figure}

To explain 
this behaviour the contributions to $\tilde{x}_{c}$ from the principal and counter jet to the centroid are separated for $\theta_{\textrm{obs}}=20^{\circ}$ in figure \ref{fig:jetcentroidscontr} (orange and green dashed lines respectively). 
The solid, blue line shows the overall centroid evolution for comparison. Initially, the contribution of the principal jet dominates in both jet models. In the non-spreading case (left panels), when the principal-jet decelerates sufficiently the counter-jet contribution becomes relevant and the centroid motion reverses. While at early times the jet core dominates the centroid calculation, as it decelerates and becomes less bright the contribution of the slower jet wings becomes more significant. At very late times $T\sim1200$ days the principal jet centroid also reverses its motion as the jet decelerates.

Two factors contribute to the earlier onset of centroid motion reversal in the laterally spreading case. As in the non-spreading case, the motion of the counter jet contributes to the reversal and this happens at much earlier times because the luminosity of the principle jet decays faster. In addition, the edges of the jet expand to $\theta_{j}>\theta_{\rm obs}$ and part of the jet begins to move backwards in the imaging plane. Consequently, the expansion becomes more isotropic, which slows down the principal jet centroid displacement and eventually also contributes to the reversal. 

\begin{figure} 
\centering
\includegraphics[width=0.9\columnwidth, height=0.3\textheight]{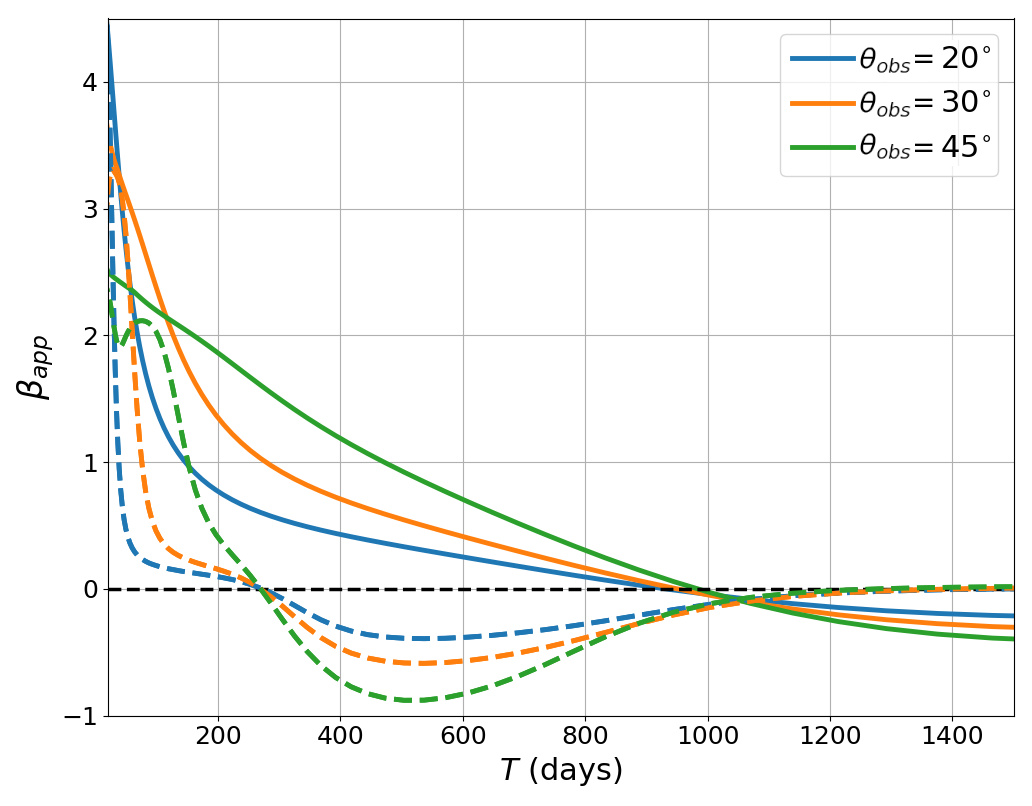}
\caption{Apparent velocity of the centroid in units of $c$. Solid (dashed) lines correspond to non-spreading (spreading) case.}
\label{fig:jetbeta}
\end{figure}

In figure \ref{fig:jetbeta} apparent velocity of the centroid in units of $c$, $\beta_{\rm app}$, is shown as a function of $T$. For all cases the centroid displacement is initially superluminal, $\beta_{{\rm app},c}>1$. The apparent velocity presents a much steeper decrease in the laterally spreading cases. The threshold $\beta_{\rm app}<0$ corresponds to the reversal of the centroid motion, which in all cases peaks at subluminal velocities. After becoming negative, the velocity asymptotically approaches $\beta_{\rm app} \rightarrow 0$ as both the principal and the counter jet decelerate (this is not shown in the figure for the non-spreading case). 

Since light curves peak when the core becomes visible to the observer $\Gamma(\theta_{\textrm{obs}}-\theta_{\textrm{c}})\sim 1$, the apparent velocity of the centroid at the peak time of the light curve can be used to estimate $\Delta \theta=\theta_{\rm obs}-\theta_{c}$ (e.g. \citealt{Mooley_2018,nakar_piran_2020}). 
A point source moving with a Lorentz factor $\Gamma$ at an angle $\Delta \theta$ with respect to the LOS has the maximum apparent velocity $\beta_{\textrm{app}}=\beta\Gamma=\sqrt{\Gamma^{2}-1}$ for an angle $\sin{\Delta \theta}=\Gamma^{-1}$. 
If the apparent velocity is obtained around the peak time, the angle $\Delta\theta$ can be estimated as $\sin{\Delta \theta}=(1+\beta_{\textrm{app}}^{2})^{-1/2}$. 

At the peak time of the light curve $T_p$, the simple estimate of the apparent velocity  $\beta_{\rm app}=\sin^{-2}{\Delta \theta}-1 \approx 3.3 , 2.0$ and $1.1$ for $\theta_{\rm obs}=20^{\circ},\ 30^{\circ}$ and $45^{\circ}$ respectively, where we have used the values of $\Delta \theta$ set by the simulations. 
The values of the $\beta_{\rm app}$ obtained from the centroid shift and the inferred $\Delta \theta$ are shown in table \ref{tab:table5}. 
The values of the numerical $\beta_{\rm app}$ are larger than those expected from the simple estimate, implying that $\Delta \theta$ is slightly underestimated in this method. The errors are larger in the spreading case. 

\begin{table}
\centering
\begin{tabular}{llll}

$\theta_{\rm obs}$ &$\beta_{\rm app}$ (coll./spread.) & Inferred $\Delta \theta$ (coll./spread.) & $\%$ error  \\ \hline
$20^{\circ}$ &\ \ \ \  \ \   $3.5/4.2$      &\ \ \ \ \ \ \ \ \  $16^{\circ}/14^{\circ}$ &  $6\%$ / $19\%$           \\
$30^{\circ}$ &\ \ \ \  \ \  $2.4/2.7$     &\ \ \ \ \ \ \ \ \  $22^{\circ}/21^{\circ}$ &  $15\%$ / $22\%$          \\ 
$45^{\circ}$ &\ \ \  \  \ \  $1.5/1.6$    &\ \ \ \ \ \ \ \ \  $33^{\circ}/32^{\circ}$ &  $21\%$ / $24\%$          \\ \hline
\end{tabular}
\caption{Observing angles inferred from the apparent velocity of the radio image centroid at light curve peak time.  }\label{tab:table5}
\end{table}

\subsection{Image structure}

The centroid is a robust characteristic of jet images which can be relatively easily obtained from observations. For brighter jets it might be possible to carry out more detailed analysis to obtain other properties \citep{Zrake_2018}. Figure \ref{fig:vertical_extent} shows the full width at half maximum (FWHM) along the $\tilde{y}$-direction, as measured at $\tilde{x}=\tilde{x}_{c}$. The case of the non-spreading jet presents only modest growth even at late times as the expansion of the jet is preferably along $\tilde{x}$. As lateral spreading leads to significantly enhanced growth in $\tilde{y}$, the FWHM grows much faster, roughly as the physical size of the image in the sky. The centroid shift and the evolution of the FWHM can confirm or give a constraint on the lateral expansion law of jets. 

\begin{figure} 
\centering
\includegraphics[width=0.9\columnwidth, height=0.6\textheight]{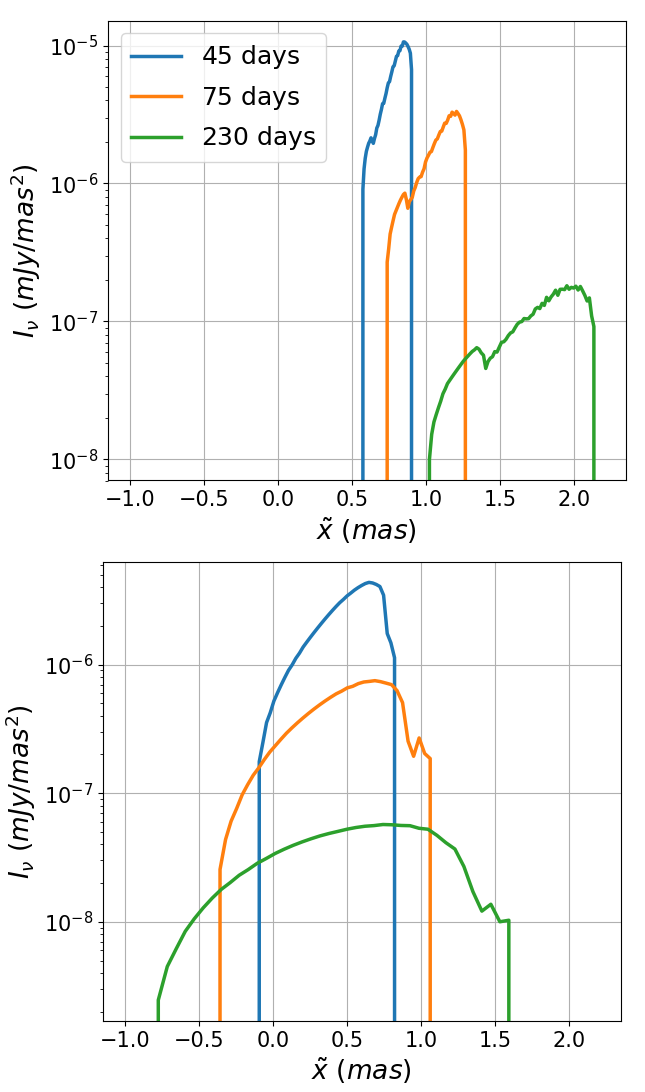}
\caption{Lateral $\tilde{y}$-averaged brightness distributions for $T=44,\ 75$ and $230$ days (blue, orange and green). Top panel: non-spreading case. Bottom panel: laterally spreading case. The distributions were obtained for $\theta_{\rm obs}=20^{\circ}$.}
\label{fig:mean_dists}
\end{figure}

\begin{figure} 
\centering
\includegraphics[width=0.9\columnwidth, height=0.3\textheight]{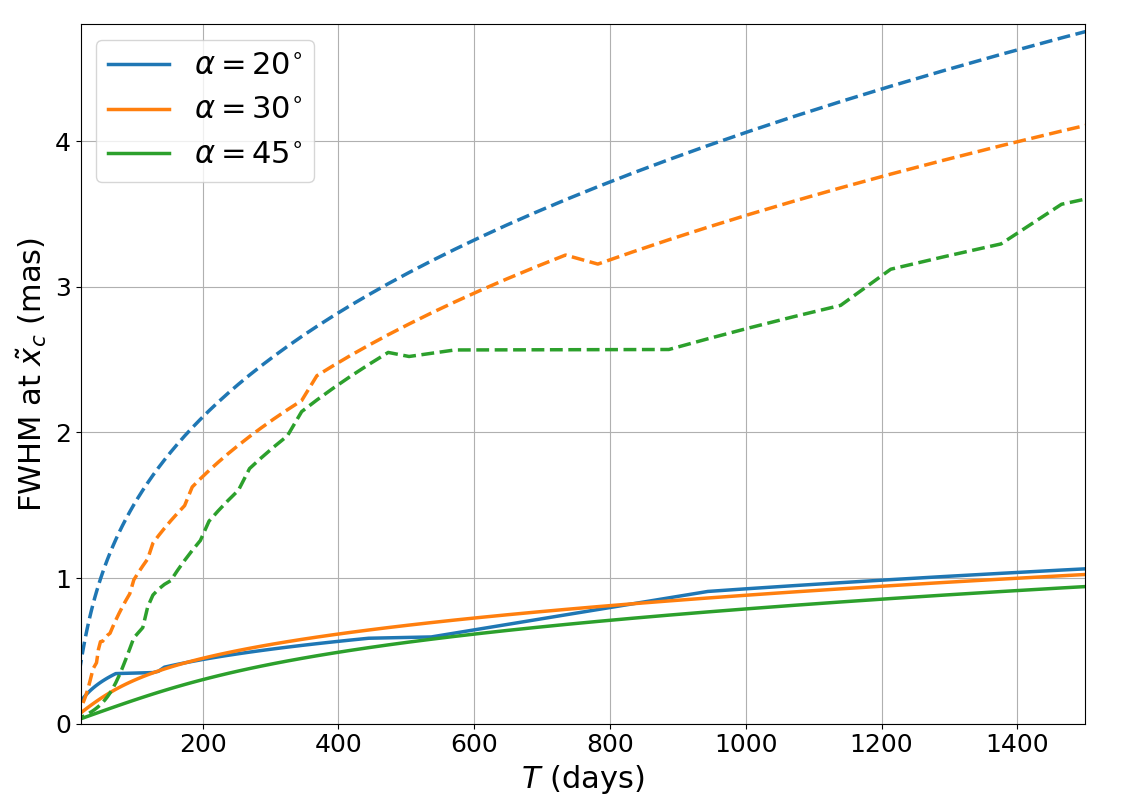}
\caption{Vertical extent of the radio images, defined as the full width at half maximum (FWHM) in the $\tilde{y}$-direction at $\tilde{x}=\tilde{x}_{c}$. The solid lines are for a non-spreading jet, the dashed lines for a laterally spreading jet. }
\label{fig:vertical_extent}
\end{figure}

Figure \ref{fig:mean_dists} shows the surface brightness distributions along the jet axis,
\begin{eqnarray}
I_{\nu,{\rm mean}}(\tilde{x}) = \frac{\int I_{\nu}(\tilde{x}, \tilde{y}) d\tilde{y}}{\Delta \tilde{y}},
\end{eqnarray}
at times $T=44,\ 75$ and $230$ days after the explosion, for $\theta_{\textrm{obs}}=20^{\circ}$. A brightness threshold $I_{\nu}/I_{\nu,{\rm max}}\ge 0.01$ was set in the images image.

The distributions in figure \ref{fig:mean_dists} trace the morphology of the full images in figure \ref{fig:images20}. For the non-spreading case (top panel), the distributions move to the right (larger $\tilde{x}_{c}$) and they become slightly wider as time passes. 
For the spreading case (bottom panel),  the spreading of the distribution is more significant compared to the propagation.  The rear edge moves backward as time passes.  The peak of the distribution is located slightly behind the front edge, before a sharp dip due to the ring which encloses the central jet material. 
We find similar results for larger inclinations $\theta_{\rm obs}=30^{\circ},\ 45^{\circ}$. These results indicate that the lateral spreading introduces distinguishable features in the brightness distributions from the non-spreading case. In particular, if the reverse motion of the rear edge is observed this would be a signature of the lateral expansion. For bright events, the centroid shift and FWHM evolution measurements might confirm or give a tight constraint on the lateral expansion law of merger jets.

\section{Applications of the spreading jet semi-analytic model to GW 170817}\label{sec:degendis}

Afterglow light curves have served as powerful tools to understand the jet physics. However, \cite{nakar_piran_2020} has discussed in detail (see also \citealt{Mooley_2018, Ryan_2020}), 
there is degeneracy among the system parameters. 
A given light curve can be compatible with a wide range of models, with varying model parameters. Light curves, especially the widths of the light curve peaks, can constrain only the angular ratio $\theta_{\rm obs}/\theta_{c}$, and each angle is not determined separately.



\begin{figure} 
\centering
\includegraphics[width=0.95\columnwidth, height=0.3\textheight]{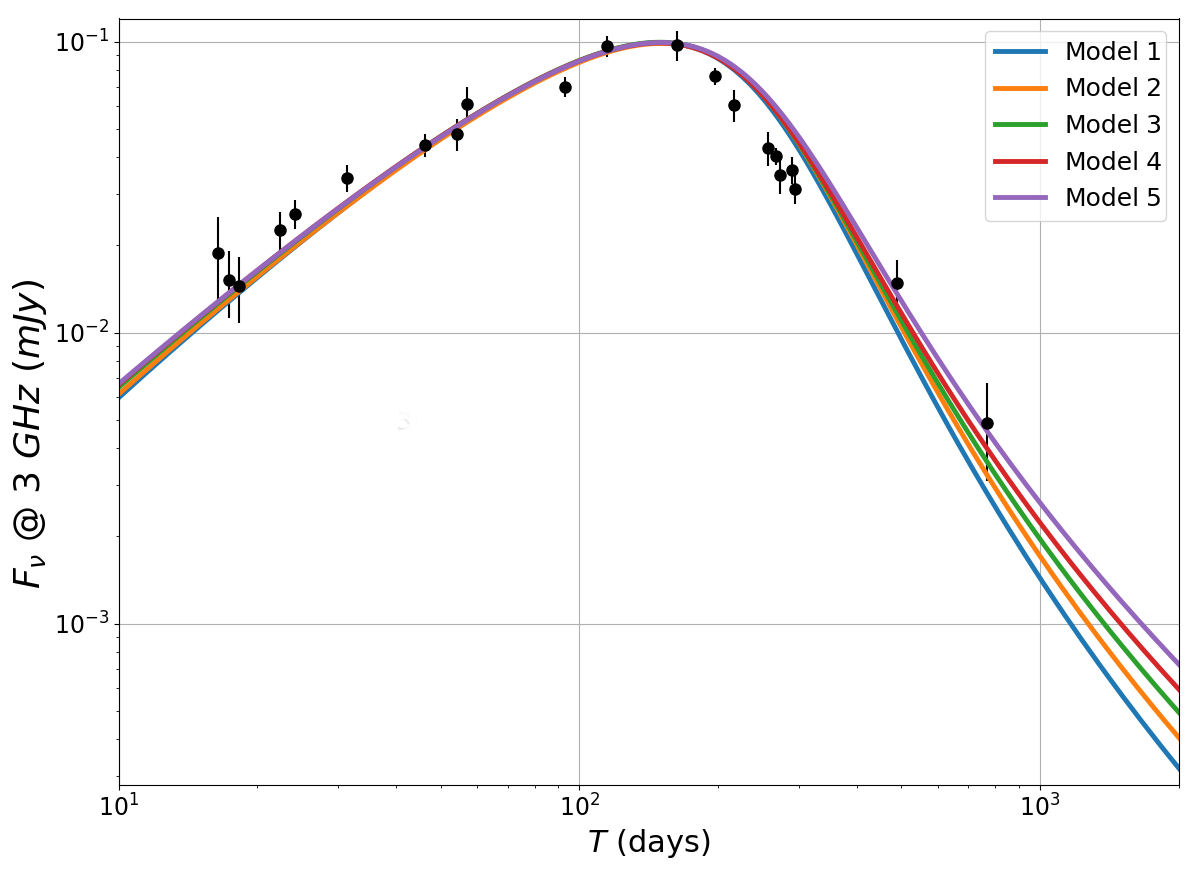}
\caption{Synthetic light curves for GRB\,170817A-like systems obtained using the parameters in table \ref{tab:datatable3}. Data points for $3$ GHz VLA observations (from \citealt{Makhathini_2020}) are shown as black circles. The spreading jet model has been assumed.}
\label{fig:degenLC}
\end{figure}

To illustrate the degeneracy, five different Gaussian jets are considered, with parameters given in table \ref{tab:datatable3}. These models were obtained by fixing the ratio $\theta_{\rm obs}/\theta_{c}$, and scaling $\theta_{\rm obs},\ \theta_{c},\ n$ and $\epsilon_{B}$ to keep $T_{p}$ and $F_{\nu}(T_{p})$ constant. We have fixed the energy $E$ while varying the CBM density $n$. 
The models span two orders of magnitude of $n$ (or equivalently a factor of $10^{2/3}$ in $l$) and one order of magnitude for $\epsilon_{B}$. The models chosen to roughly match radio ($\nu = 3\ GHz$) data for the afterglow of the neutron star merger GW170817. 
The light curves are shown in figure \ref{fig:degenLC} in which the light curves are identical around the peak.

\begin{table}
\centering



\begin{tabular}{l|lllll}

 & $\ n\ (\textrm{cm}^{-3})$ & $\ \epsilon_{B}$ & $\theta_{c}$ (rad) & $\theta_{j,0}$ (rad) & $\theta_{\rm obs}$ (rad) \\ \hline
Model 1 & $ \ 1.7\times 10^{-5}$    &  $0.04$ & $\ \ 0.05$        & $\ \ \ 0.16$       & $\ \ \ \  0.23$  \\ 
Model 2 & $ \ 4.8\times 10^{-5}$    &  $0.02$ & $\ \ 0.06$        & $\ \ \ 0.19$       & $\ \ \  \ 0.28$  \\ 
Model 3 & $ \ 1.7\times 10^{-4}$    &  $0.008$ & $\ \ 0.07$        & $\ \ \ 0.23$       & $\ \ \  \ 0.33$  \\ 
Model 4 & $ \ 4.8\times 10^{-4}$    &  $0.004$ & $\ \ 0.08$        & $\ \ \ 0.26$       & $\ \ \ \  0.37$   \\ 
 Model 5 & $ \ 1.2\times 10^{-3}$    & $0.003$  & $\ \ 0.09$        & $\ \ \ 0.30$       & $\ \ \ \  0.42$  \\ 
\end{tabular}
\caption{Simulation parameters for the jets in \ref{sec:degendis}. The other jet parameters are fixed, $E = 4\pi\epsilon_{c}= 10^{52.4} \textrm{ergs}$, $\Gamma_{c}=300$, $\epsilon_{e}=10^{-1.4}$, $p=2.16$, $D_{L}=41.3$ Mpc.}\label{tab:datatable3}
\end{table}

\begin{figure} 
\centering
\includegraphics[width=0.9\columnwidth,height=0.3\textheight]{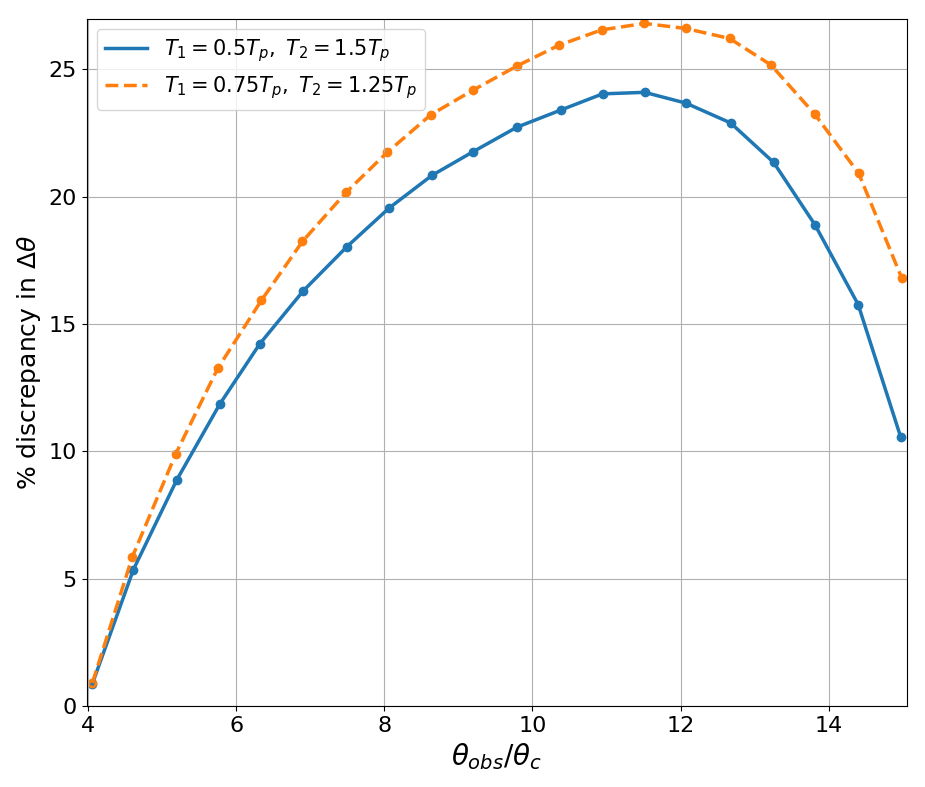}
\caption{Error in the estimation of $\theta_{\rm obs}$ from synthetic images as a function of $\theta_{\rm obs}/\theta_{c}$ for two different observing windows. }
\label{fig:thetestimate_err}
\end{figure}

\begin{figure*} 
\centering
\includegraphics[width=1.8\columnwidth, height=0.85\textheight]{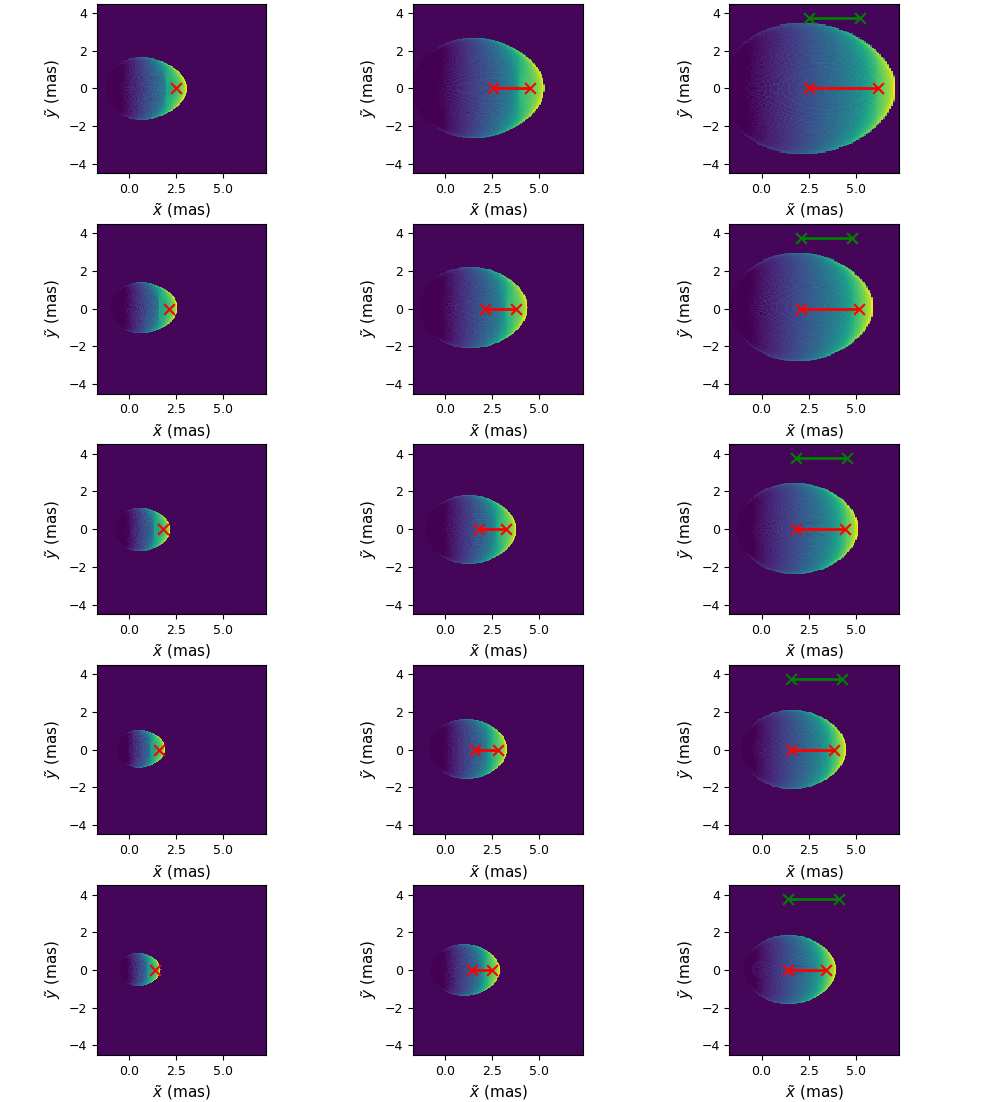}
\caption{Synthetic radio images ($\nu=4.5 GHz$) corresponding to the degenerate light curves in figure \ref{fig:degenLC}. The rows correspond to models $1-5$ respectively. The left, central and right columns correspond to times $T=75,\ 150$ and $230$ days. The centroid positions are shown with red crosses, and the centroid displacment with respect to $T=75$ days are shown with red lines. The centroid displacement reported in \citep{Mooley_2018} is shown with green lines. The surface brightness in each image is normalized to $I_{\nu,{\rm max}}$, with the colour map covering the range $0.01 I_{\nu,{\rm max}}-I_{\nu,{\rm max}}$}
\label{fig:degenIM}
\end{figure*}

The properties of radio images depend directly on the geometry and dynamics of the shock. By combining the constraints from the light curves and images we can distinguish different models (\textbf{e.g. \citealt{Mooley_2018, Ghirlanda_2019, Ryan_2020})}. Figure \ref{fig:degenIM} shows radio images obtained for models $1-5$ at 75 and 230 days (the days for which VLBI images for GW170818 were reported in Mooely et. al 2018), and around the peak time at 150 days.
For these cases, the model images also present a brightness distribution which decreases gradually from front to back.
Note that the morphology is the same for all models, up to scaling of the image size.
This is due to the ratios $\theta_{\rm obs}/\theta_{j}=1.4$ and $\theta_{\rm obs}/\theta_{c}=4.6$ having been fixed (the angle ratios have been determined by the peak of the light curve). 
The position of the centroid in each image (red crosses) and the displacement with respect to the position at $75$ days (solid red lines) is also shown. 
At $75$ days the centroid calculation is dominated by the leading bright point in the image and as lateral spreading slows the radial expansion of the wings of the shell, the centroid falls behind the the front of the image, as seen in the images for $150$ and $230$ days post-merger. 
The values of the centroid displacement between $T=75$ days and $T=230$ days, and the apparent velocity computed as $\beta_{\rm app}=\Delta x_{c}/c \Delta T$, are reported in table \ref{tab:resultstable}. 
$\Delta \theta$ is also computed from $\beta_{\rm app}$ using the point particle approximation as $\sin{\Delta \theta}=(1+\beta^{2}_{\rm app})^{-1/2}$. 
In this case the approximation results provide a much better estimate of the the inclination, with deviations of the order of a few percent for rather small $\theta_{\ obs}/\theta_{c}$. The errors in the $\Delta \theta$ estimates were larger for the parameter sets discussed in section 3.1 (i.e. table \ref{tab:resultstable}, $19-24\%$ for the laterally spreading jets), in which $\Delta \theta$ was obtained by using the point-emitter approximation. We now consider model 3 in \ref{tab:datatable3}, and by changing the viewing angle $\theta_{\rm obs}$ (the other parameters including
$\theta_{c}$ are fixed), we reevaluate errors as a function of $\theta_{\rm obs}/\theta_{c}$. The results are shown in \ref{fig:thetestimate_err} where the apparent velocity $\Delta \tilde{x}_{c}/\Delta T$ have been estimated for two time windows $\Delta T=T_{2}-T_{1}=0.5 T_{p}$ or $T_{p}$. We find that the errors (or the discrepancy) are larger for larger angle ratios. However, merger afterglow observations (i.e. detailed light curves and radio images) will be available only for bright events. The detail afterglow
modeling is likely to be carried out only for low inclination events, for which the point-emitter estimates are accurate, and the apparent velocity estimates are not sensitive to the choice of the time window. 

The observed centroid displacement obtained from VLBI imaging between $75$ and $230$ days post-merger, as reported in \citep{Mooley_2018}, is shown for comparison (green solid lines delimited by crosses). 
They found a centroid displacement of $2.7\pm 0.3$ mas, equivalent to an apparent velocity $\beta_{\rm app}=4.1 \pm 0.5$. 
Figure \ref{fig:degenIM} shows how radio imaging can be used for parameter inference. 
While the details of the morphology of the images is the same for all models, the evolution of the centroid is determined by growth of the emitting region, which in turn is related to how fast the shock can expand. 
For example, model $1$ ($5$) involves a slowly (rapidly) expanding jet as the density is the lowest (highest) in the set.  
Visual inspection shows that model $1$ ($5$) results in excess (too little) centroid displacement to be compatible with observations. 
Models $2,\ 3$ and $4$ provide a centroid displacement in agreement, within error, of the observational value. Note that while this analysis can provide a constraint on the viewing angle, the uncertainty is dominated by the model uncertainties (the prescribed jet structure in the semi-analytic model) and the data (the distance to GW170817 and VLBI uncertainties).

The direct comparison of between the observed and numerical centroid shift can provide a constraint on $\theta_{\rm obs}$ without using the approximation $\Delta \theta \sim \beta_{\rm app}^{-1}$ (a similar analysis has been presented in \cite{Mooley_2018} using results from hydrodynamics simulations).

\begin{figure} 
\centering
\includegraphics[width=0.9\columnwidth,height=0.3\textheight]{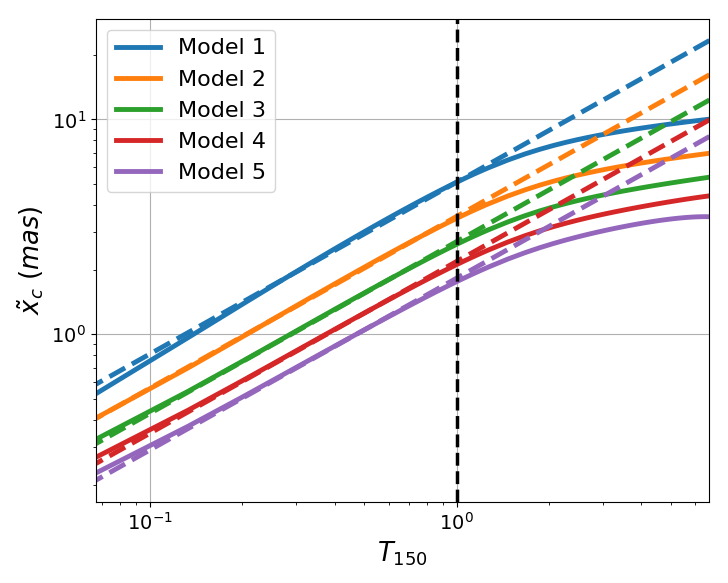}
\caption{Centroids of the images for models $1-5$. The dashed lines are for power-law fits $\tilde{x}_{c}\propto t^{0.86}$. The color scheme is the same as in figure \ref{fig:degenIM}.} 
\label{fig:degenCENT}
\end{figure}

The two methods described above make it possible to break the degeneracy in the light curve models using two radio images (e.g., \cite{Mooley_2018}). 
Figure \ref{fig:degenCENT} shows the centroid evolution for each of the five models where the $\tilde{x}_{c}-T$ curves do not overlap as long as the centroid does not move backwards. 
In all cases the curves behave as broken power-laws $\tilde{x}_{c} \propto T^{\omega}$, where it is found that $\omega \approx 0.86$ provides a good fit until the light curve peak time $T\sim 150$ days (dashed lines in the figure), and the curve is flattened as $\omega\approx 0.4$ since after the peak the emission is dominated by the core radiation, assuming the simple spreading jet $\Gamma \sim T^{-1/2}$, the centroid shift can be approximated as $\tilde{x}_{c}\sim \beta_{\rm app}T  \sim T^{1/2}$. Since the jet wing propagate slower, the small contribution might make the scaling index smaller in the numerical model.

\begin{table}
\centering
\begin{tabular}{l|llll}
& $\Delta x_{c}$ (mas) & $\beta_{\rm app}$ &$\Delta \theta$ (rad) & $\%$ error \\ \hline
Model 1 & $\ \ \ \  3.7$ & $\ 5.6$ &\ \ \ \ $0.18$ &    $\ 3.5 \%$    \\ 
Model 2 & $\ \ \ \  3.1$ & $\ 4.7$ &\ \ \ \ $0.21$ &    $\ 4.6 \%$    \\ 
Model 3 & $\ \ \ \ 2.6$ & $\ 4.0$ &\ \ \ \  $0.24$ &    $\ 4.5 \%$    \\ 
Model 4 & $\ \ \ \ 2.3$ & $\ 3.5$   &\ \ \ \ $0.28$ &   $\ 5.0\%$    \\ 
Model 5 & $\ \ \ \ 2.0$ & $\ 3.1$&\ \ \ \   $0.31$  &   $\ 5.3\%$    \\        
\end{tabular}
\caption{Observable parameters extracted from the synthetic radio images shown in \ref{fig:degenIM}. }\label{tab:resultstable}
\end{table}

\section{Conclusions}\label{sec:conclusions}

We have presented radio image calculations from semi-analytic modeling, focusing on comparison between spreading and non-spreading approximations. We have also shown a comparison of these models to the analytic point source approximation and a comparison of the spreading jet semi-analytic model to the observations of GW170817.

While spreading is not immediately apparent in the rising phase of afterglow light curves, it affects the slope and peak time, and can certainly affect images.  We find that lateral spreading has an important effect on the morphology of the images and the evolution of the centroid even at early times. It is crucial to include proper lateral spreading effects when analyzing radio images of neutron star merger jets at late times.

If neutron star mergers happen in a higher density or/and higher pressure region compared to that where usual short GRBs have been detected 
(e.g., the disks of active galactic nuclei; \citealt{perna21}), the lateral expansion law of a jet might deviate from a simple jet hydrodynamic model. Jet radio images  might be able to reveal such confinement effects, although observations of such events are likely to be very challenging due to the distance to the sources and/or the background noises.

We study two methods to determine the viewing angle from radio images: the point-emitter approach $\Delta \theta \sim \beta_{\rm app}^{-1}$ (e.g. \citealt{rees_1966, Mooley_2018}) and direct comparison of two or more images (e.g. \citealt{Mooley_2018})
by using a sample of Gaussian structured jets with parameters that can roughly explain the light curve of the GW170817 afterglow. The direct comparison of the centroid displacements gives the viewing angle to be $\theta_{\rm obs}\approx 0.32$, which is consistent with previous studies (e.g. \citealt{Mooley_2018, Ghirlanda_2019, Hotokezaka_2018, Lamb_2019_3, Troja_2019a}).
We find the simpler point-emitter approximation is
accurate especially for the GW 170817 case.



\section*{Acknowledgements}

We acknowledge the referee Ehud Nakar for his valuable and constructive suggestions and Om Salafia for useful discussions.
This research was supported by STFC grants and a LJMU scholarship. GPL is supported by the STFC via grant ST/S000453/1.

\section*{Data availability}
The data underlying this article and additional plots will be shared on reasonable request to the corresponding author.

\bsp	
\label{lastpage}
\end{document}